\documentclass[12pt, preprint]{aastex}

\newcommand{\hstar}{h_\star}
\newcommand{\hsun}{h_\odot}

\shorttitle{Duty cycle of asteroseismic
observations} \shortauthors{Mosser \& Aristidi}

\begin{document}

\title{Duty cycle of Doppler ground-based asteroseismic observations}

\author{Mosser Beno{\^\i}t\altaffilmark{1}}
\affil{LESIA, Observatoire de Paris/CNRS UMR 8109, F-92195 Meudon, France}
\email{benoit.mosser@obspm.fr}

\and

\author{Aristidi Eric\altaffilmark{2}}
\affil{LUAN, Universit\'e de Nice Sophia Antipolis, Parc Valrose, F-06108 Nice cedex 2, France}
\email{aristidi@unice.fr}
\altaffiltext{1}{Research associate at IAP, CNRS}

\altaffiltext{2}{Second winterover at the French-Italian Concordia Station, Dome C, Antarctica}

\begin{abstract}
We report the observations of the clear sky fraction at the Concordia station during winter 2006, and derive
from it the duty cycle for astronomical observations.
%Time series for asteroseismology must be as continuous as possible, over a duration as long as possible, as is achieved by SoHO or the networks dedicated to helioseismology, and as will be achieved by the CoRoT mission for asteroseismology. Therefore, the positive ground-based asteroseismic observations solar-like oscillations, mostly obtained with single-site measurements, must now be followed by long duration multi-siteobservations. 
Performance in duty cycle and observation duration promotes Dome C for efficient asteroseismic observations. This performance is 
analyzed and compared to  network
observation.  For network observations, simulations
were run considering the helioseismic network GONG as a
reference.\\ Observations with 1 site in Antarctica provide
performance similar or better than with a 6-site network, since
the duty cycle limited by meteorology is as high as 92\%. On
bright targets, a 100-day long time series with a duty cycle about
87\% can be observed, what network observation cannot.
\end{abstract}


\section{Introduction}

Stellar evolution theory, central to most areas of modern
astrophysics, is not sufficiently tested by observations. The
analysis of stellar oscillation modes, which propagate inside
stars down to the core, constitutes a powerful tool to probe their internal structure, and provides the necessary observational basis for the theory. Already applied to the Sun with a remarkable
success, after pioneering observations at the South Pole by Grec, Fossat \& Pomerantz (1980), this technique is now opening up to stars showing solar-like oscillations, both from space and from the ground.

Successful monosite Doppler asteroseismic observations were performed
recently with \'echelle spectrometers (CORALIE at the Swiss 1.2-m
telescope in Chile, HARPS at the 3.6-m ESO telescope, UCLES at AAT
in Australia). Mostly sub-giants were observed. Only global
oscillations parameters, as the large splitting, were measured,
eigenfrequencies being identified only for the brightest targets.
The window effect due to daily interruptions leads to large
uncertainties and ambiguities in the frequency determination, and
the frequency resolution remains limited by short observations.
Two-site observations were performed only on bright targets, with
a time series with efficient duty cycle limited to about a few
days: $\alpha$ Cen A and B were observed with UCLES/AAT and
UVES/VLT, (Bedding et al. 2004); $\nu$ Indi was observed with
UCLES/AAT and CORALIE/Euler telescope (Bedding et al. 2006).

The space project CoRoT, to be launched in December 2006, will
provide high-precision photometric observations, with an
unprecedentedly high duty cycle (Baglin et al. 2002). After
CoRoT observations, we believe that the major contribution of
ground-based observation will consist in monitoring the Doppler
signature of stellar oscillations. In fact, even in an excellent site as Dome C, photometric measurements remain unable to provide the necessary precision. 

In fact, spectrometric observations will be able to detect
$\ell=3$ oscillation modes that cannot be analyzed in photometry,
and will be less affected by stellar activity noise, increasing
the complementarity with space-borne photometric observations (Toutain et al. 1997). Such observations need long duration, in order to obtain a very
precise determination of the eigenfrequency pattern. Precision in the
frequency measurement $\delta\nu$ varies as the square root of the
observation total duration $T$
%$$\delta\nu^2\ = \ f(SNR) \ {\Gamma \over T}$$ where $f(SNR)$ is a function of the signal-to-noise ratio, equal to unity at high SNR, and $\Gamma$ is the FWHM linewidth
(Libbrecht 1992). Precise eigenfrequencies translate linearly into
precise modelling, according to
inversion technique. %a frequency precision improved by a factor or 2, obtained with, crudely speaking, an observation multiplied by 2, reduces the uncertainty in the modelling by the same factor.

In this paper, we address the question of optimized duty
cycle of ground-based observations. Possible solutions are
presented in Section 2, in the case of long-duration observations. Meteorological conditions and maximum
duration of efficient observation are examined in Section 3. We
report the observations of clear sky fraction made during the
winterover of Eric Aristidi at the Concordia station. In Section
4, we discuss the properties of spectroscopic asteroseismic observations at Dome C, and compare them to what could be obtained from network observation. The parameters for the comparison correspond to the observations of bright solar-like stars, as proposed by the observational projects SONG (Grundahl et al. 2006) or SIAMOIS (Mosser et al. 2006).
The duty cycle of the time series and the energy in
the first diurnal alias of its Fourier spectrum are calculated as
a function of the total observation duration, for different
observation conditions.

\section{How to obtain ground-based continuous observations?}

\subsection{Network}

The many results obtained by the helioseismology networks have
shown their capability to provide long duration time series with a
high duty cycle, better than 80\% when averaged over long
durations and about 90\% for short runs, as reported by the GONG
project (\texttt{http://gong.nso.edu}, Harvey et al. 1996). Such
performance was prepared by intensive efforts in site survey (Hill
et al. 1994). The duty cycle obtained by the WET campaigns
dedicated to stellar variability photometric measurements has
never reached such values, despite a greater number of sites
involved in the network, but without a strict selection of the
sites involved in the network. Dolez et al. (2006) report for
example a 18-day campaign with a duty cycle of 66.5\%.
We note that the baselines of the SONG network project,
dedicated to the Doppler monitoring of solar-like oscillations in bright stars, are clearly derived from the
organization of the GONG project. The concept of SONG involves
small size telescopes and iodine-cell \'echelle spectrographs
(\texttt{http://astro.phys.au.dk/SONG/}, Grundahl et al. 2006), to be
installed in selected sites similar to the ones of GONG. Accordingly, and taking into account the excellent duty cycle performance achieved by the network GONG, we consider in this  paper its characteristics as a reference for a network dedicated to Doppler measurements.

\subsection{Antarctica}

In this paper, we do not consider possible observation at the
South Pole, where strong catabatic winds yield recurrent cloudy
periods, for a duty cycle limited to about 50\% during winter. Attention is
focussed on possible asteroseismic observation to be conducted at
Concordia station, located on the Dome C summit on the Antarctic
plateau (75$^\circ$6' South, 121$^\circ$21' East, altitude of 3250
m). Intensive site testing is currently made there (Aristidi et
al. 2005 and references therein). The site presents very positive
characteristic features for astronomical observation: very low
temperature, humidity, very limited winds (Aristidi et al. 2005).
The whole atmosphere nighttime seeing was first observed during
the 2005 winter (Agabi et al. 2006). Kenyon et al. (2006) report a low scintillation level limited to about 50 $\mu$mag. The
amplitudes of solar-like oscillations, about a few ppm, remain too tiny for photometric asteroseismologic ground-based observations. In that case, spectrometric observations are definitely required.

Currently, observing conditions at Dome C are governed by the
limited available infrastructures: in a near future, only light
operations with small telescopes can be achieved. The SIAMOIS
project (Mosser et al. 2006) has been designed accordingly. The
SIAMOIS concept is based on Fourier Transform interferometry,
which leads to a small instrument developed for the harsh
conditions in Antarctica. Applied to the Sun in the 1980's,
Fourier Tachometry was chosen for the GONG helioseismic network
after a long study of competing measurement strategies, and it
forms the basis of the Michelson Doppler Imager instrument on the
SOHO spacecraft as well as the Velocity and Magnetic Imager on the
forthcoming SDO spacecraft. Performance of a Fourier Tachometer
for asteroseismology was estimated in Mosser et al. (2003).

\section{Duty cycle and observation duration}

Many factors may limit the duty cycle of ground-based observation. We focus here on two major points: meteorology and sky brightness due to twilight. Other factors, such as instrument down time, are not considered.

\subsection{Clear sky fraction}

\begin{figure}
\plotone{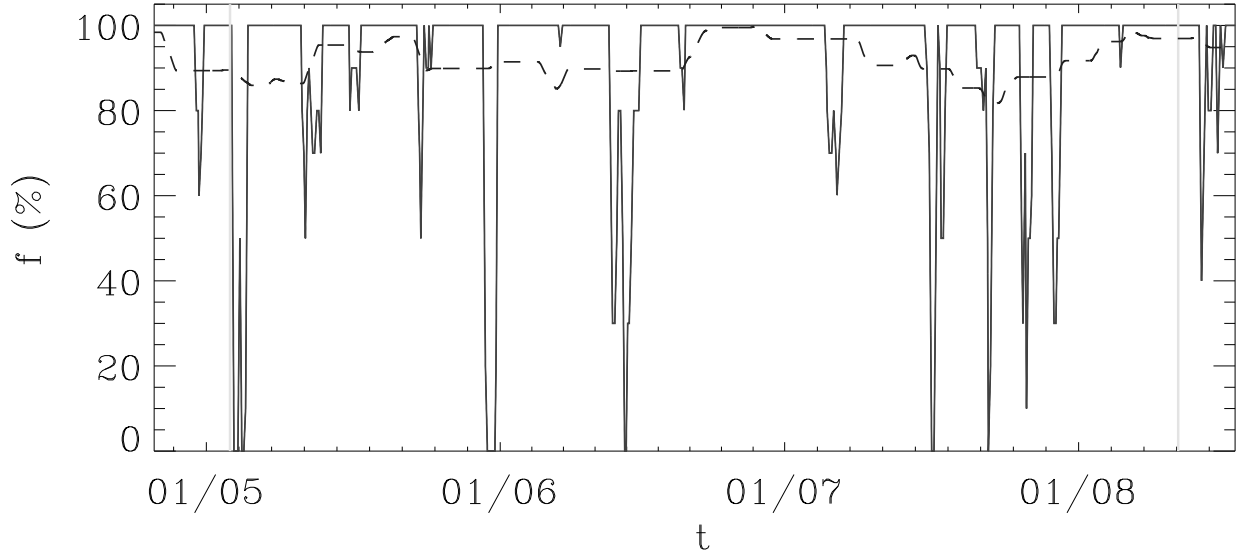}%\plotone{clearskycycle.ps}
\caption{Full line: clear sky fraction $f$ at Dome C, as observed
during the 2006 wintering. Dashed line: duty cycle at Dome C,
smoothed over a 10-day period. Lines at May 03 (embedded in a
2-day bad weather event) and August 11 represent respectively the
beginning and the end of the polar night at Dome C. The
meteorological duty cycle exceeds 92\% during 3
months.}\label{clearsky}
\end{figure}

The clear sky fraction at Dome C was observed during the 2006
winterover. It was estimated visually several times a day, using a
scale from 0 (no visibility) to 1 (cloud-free sky). We consider
here the data from March 21 to September 21 (Fig \ref{clearsky}).
Non-clear sky means the presence of thin cirrus or thicker clouds.
During this period, the clear sky fraction was greater than 0.9 for  84\% of the time. The average number of consecutive clear days (with a clear sky fraction $>0.9$) was about 6.8. 

From this global fraction $\bar f$, we have derived the condition of observation in a given direction. We simply assumed that, in that case, the observation is either possible ($f=1$) or not possible ($f=0$). An observed cloud sky fraction $\bar f$ of duration $T$ translates into an obscure window of duration limited to $\bar f T$, with possible observation during the remaining period $(1-\bar f)T$.
We have then directly obtained the window function for observations to be conducted at Dome C. The duty cycle limited by meteorology at Dome C is superimposed on the clear sky fraction in Fig.~\ref{clearsky}.

We have modelled the sky fraction of an asteroseismologic network
according to the experience and data of the GONG site survey (Hill
et al. 1994), under the assumption that the data collected for the
Sun in diurnal conditions can be adapted to night condition. We
adopted for the network configuration the one of the GONG network, with 6 sites (4 in the northern and 2 in the southern hemisphere) where the clear sky fraction was intensively measured  (Hill et al. 1994), making use of the same model for cloud apparition (Hill \& Newkirk 1985).

\subsection{Time series}

The determination of the observing time series in each site
depends first on the star altitude $h$, as a function of its
coordinates, of the latitude and longitude of each observing site,
and of the local hour angle of the star.  Possible observation in
a given site requires the star to reach a certain altitude
$\hstar$, but also the Sun to be below enough the horizon
(parameter $\hsun$). Times series were produced in many
configurations. We have limited the extent of these parameters to
plausible values accounting for realistic observation condition.

The parameter $\hsun$ is of greatest importance for observation at
Dome C, with the Sun never below $-8^\circ$ at noon, implying that
neither astronomical nor nautical twilights occur in this site.
Values within the range [$-12^\circ,\ 0^\circ$] were explored, in
order to consider different observation conditions, on faint or
bright targets. However,  future targets of ground-based
asteroseismology will clearly be very bright targets, because they
will be observed with small telescopes. Then, the contrast
with the sky background will be high enough to permit observations
during twilight, with values of $\hsun$ much smaller than the
astronomical twilight definition. Observations realized by Eric
Aristidi have shown that observation of bright ($mV<5$) targets
when the Sun is lower than $-4^\circ$ provides a sufficient
contrast between the sky and the target. Therefore, simulations
requiring the parameter $\hsun$ to be fixed were realized in the
case $\hsun = -4^\circ$.

The  $\hstar$ value must be high enough, since the decrease of
signal with airmass is always important; due to absorption, blue
photons are strongly absorbed, and due to seeing, photons do not
enter easily the fibre or the slit of the spectrometer. Simulations were realized for
airmass of 2.5, 3 or 4 ($\hstar = 23.5$, 19.5 or $14.5^\circ$ respectively). For a network of 6 sites, we have checked that the parameter $\hstar$ plays a limited role; its value
is however important when extrapolating the analysis to the case
of a network with less than 6 sites.
We have measured that, for a network, a 1$^\circ$ decrease of $\hstar$ improves the duty cycle by only 0.2\%; this proves in fact the quality of the network, since 1$^\circ$ gain in one single site at latitude 30$^\circ$ represents for an equatorial target 9.2 min more observation, hence an increase of 0.6\% duty cycle.

We included 2\% overhead into the duty cycle at Dome C, for
operations as calibration, or back rotation of the telescope. Therefore we automatically limit the maximum duty cycle at 98\% for Antarctica observations.

\subsection{Simulations}

We have constructed time series with alternation of clear and
cloudy weather, using a Monte Carlo model as described in Hill \&
Newkirk (1985), and with their meteorological parameters for the
selected sites (probability of clear sky and mean interval between
cloudy period). After having run many simulations, we have
selected a median realization, normalized to the mean observed
performance of the observing runs reported by the GONG network
(duty factor equal to about 85\%). Practically, this normalization
was achieved when considering only the condition on $\hstar$ (with
the Sun as the target), and no other condition on $\hsun$. We have
checked that, in such conditions, the simulations gave significant
and reproducible results.

We tried to limit the number of parameters of the simulation as
low as possible. Simulations have shown that seasonal effects are
limited, and therefore the right ascension influence is very
limited too. Then, for sake of simplicity, we have run the
simulations with a right ascension value $\alpha=18\mathrm{h}$,
just in order to cophase network and Antarctica observations.
Concerning the declination, we chose to focus on the most
favorable case for the network simulation: $\delta = 0^\circ$. The
variation of the duty cycle with the stellar declination is
discussed hereafter.

\begin{figure}
\plotone{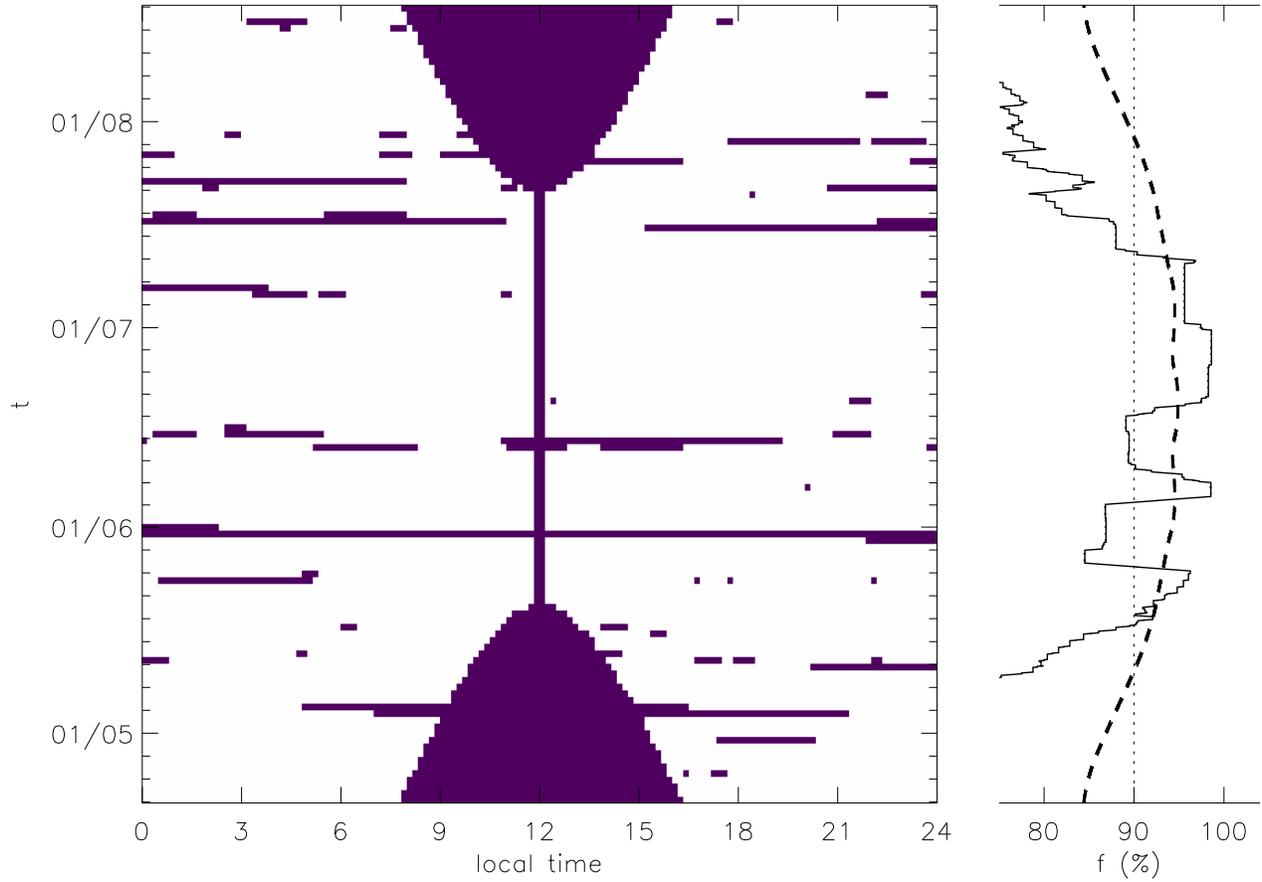}%\plotone{superfenpole.ps}
\caption{Left panel: observed window for Dome C, taking into
account the meteorological conditions recorded in 2006 and the Sun
altitude (at least $4^\circ$ below horizon). The simulation also
considers 20 minutes overhead per day, at 12:00 local time, for
calibration and back rotation of the telescope. The right panel
shows the variation with time of the duty cycle averaged over
10-day periods (full line) and the integrated duty cycle (dashed
line).
}\label{tableaufenpole}\end{figure}

\begin{figure}
\plotone{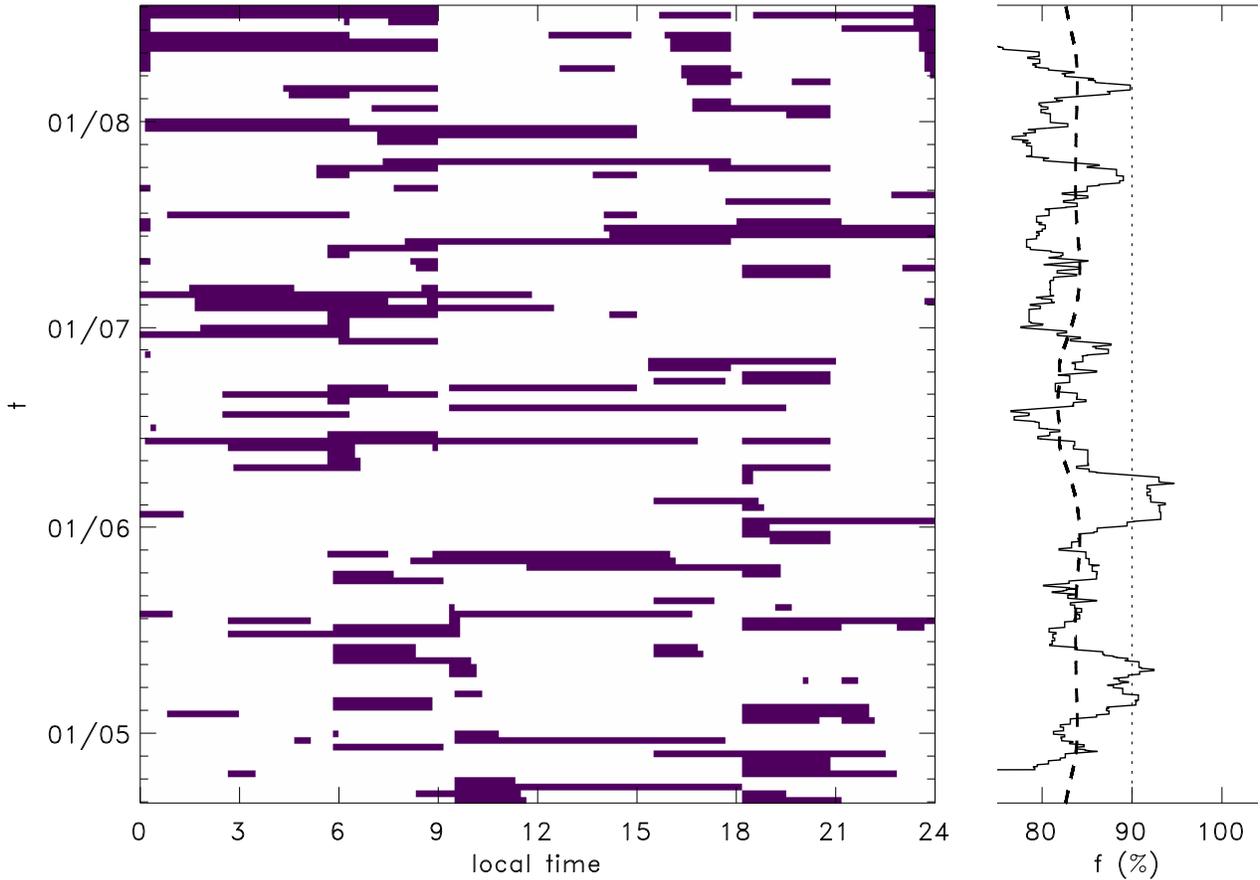}%\plotone{superfenreseau.ps}
\caption{Same as Fig.~\ref{tableaufenpole} but for the simulated
window for a 6-site network, taking into account the
meteorological conditions derived from GONG (very short
interruptions, practically related to local meteorological
conditions, are not considered). Sidereal limitations, shifting by
4 min/day, and solar limitations, stationary with local time, can
be easily distinguished. The maximum duty cycle is less  than for
Antarctica observations, but its decrease with increasing
observation duration is much less sensitive. Stellar coordinates for this simulation: $\alpha=$18:00, $\delta=0^\circ$, $\hstar = 19.5^\circ$ (airmass = 3).\label{tableaufenreseau}}\end{figure}

Figures \ref{tableaufenpole} and \ref{tableaufenreseau} present respectively the window functions observed at Dome C in 2006 or modelled for a 6-site network, represented for 140 days around the Southern summer solstice. The window function at Dome C is dominated by the sky brightness at noon local time in the days following (preceding) the beginning (end) of the night.

\section{Discussion}

\subsection{Time series}

In order to compare the simulations, we have calculated the
filling factor, averaged over 10-day periods, for a network with 4
to 6 sites and for Antarctica.

\begin{figure}
\plotone{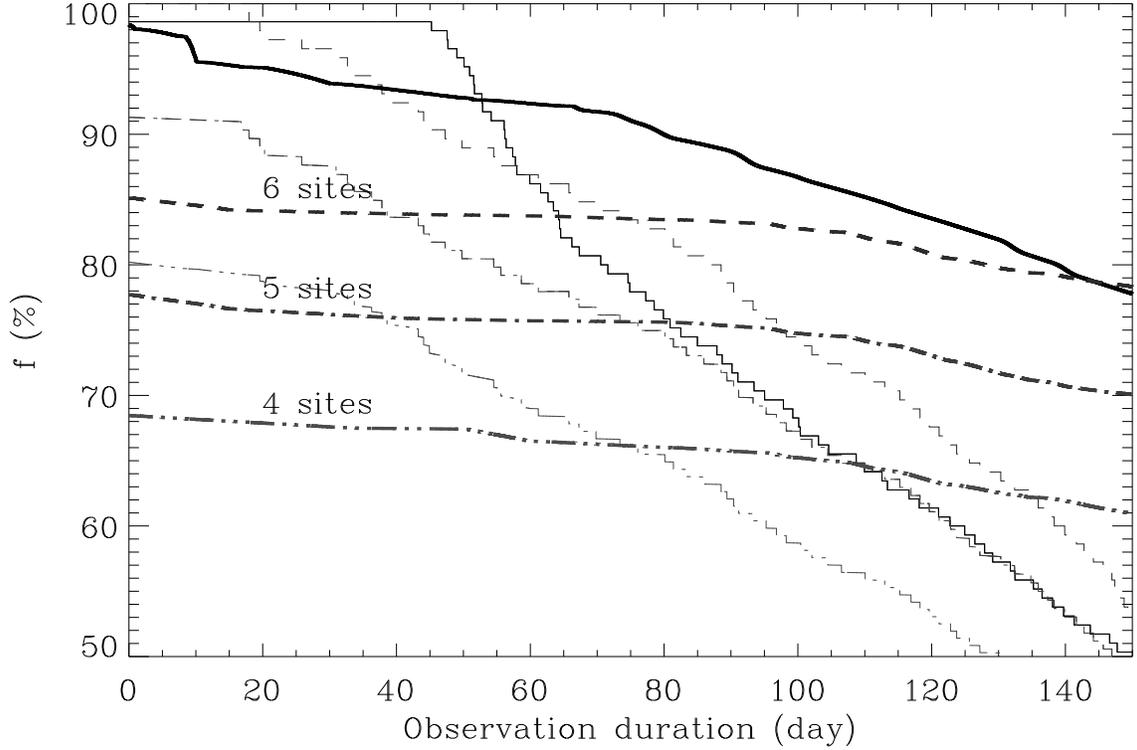}%\plotone{compar.ps}
\caption{Comparison of the daily duty cycle distributions,
at Dome C (full light line) or for a multi-site network (dashed
light line: 6 sites; dot-dashed line: 5 sites; dot-dot-dashed
line: 4 sites), and corresponding integrated duty cycle (heavy
lines). Dome C provides long period with a duty cycle about 98\%
(not 100\% in order to account for necessary diurnal operations,
such as back rotation of the telescope). With 6 sites, network
observations have a better daily duty cycle distribution only for
periods longer than about 80 days. However, the integrated duty
cycle remains better and Dome C for more than 4 months.
Simulations for 5 or 4 sites depend on the absent site(s): results
presented here have been averaged. Calculations for the network
have been considered in the favorable case of an equatorial
target, and for mean weather conditions. Simulations consider
$\hsun \le -4^\circ$ and $\hstar \ge 19.5^\circ$ (airmass = 3).}\label{compar}\end{figure}

We have sorted the results as a function of time (Fig.~\ref{compar}). Antarctica observations provide a better duty cycle for runs limited to about 2 months. For longer runs, the Sun skimming horizon appears as a limit. In that case, despite the shift between local and sidereal times, network observation gives a higher duty cycle, that nevertheless decreases sensitively with the total duration (respectively below $70$\% and $60$\% for 4 and 5-month long observations).

We have analyzed the importance of the Sun location by calculating the duty cycle obtained during the 100-day long night as a function on the condition on the altitude $\hsun$. Table~\ref{simuhsol} shows the importance of observing targets bright enough in order to be observed with high SNR even with the Sun not far below horizon. Such targets are already identified (Table~\ref{target}).
However, the time series analysis is not sufficient for complete estimations or comparisons: the signature in the Fourier spectrum of the data interruption must be checked.

\subsection{Frequency analysis}

\begin{figure}
\plotone{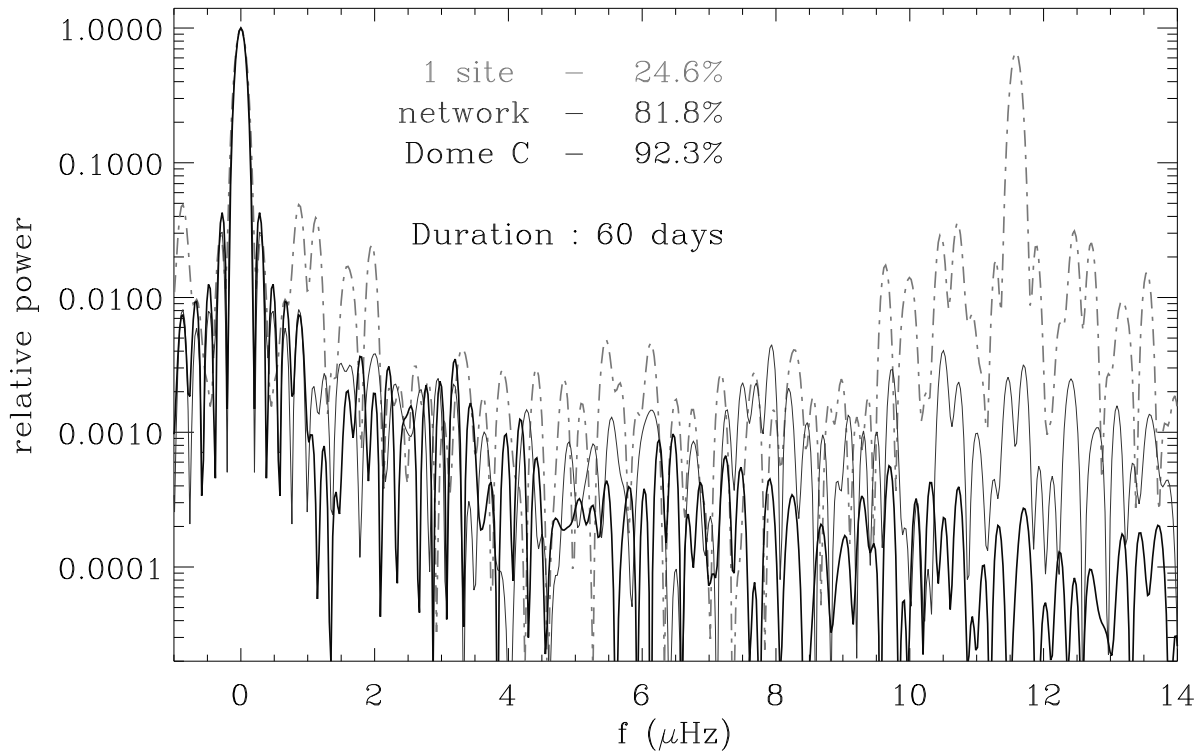}%\plotone{compar60.ps}
\caption{Window function obtained at Dome C during 60 days, compared to simulations for network and single-site observations (respectively grey and dotted lines). There is 10 times less energy emerging around the first day alias at 11.57 $\mu$Hz for Antarctica observations than for a network. The gain compared to the single site case is higher than 1000.
Simulations consider $\hsun \le -4^\circ$ and $\hstar \ge 19.5^\circ$ (airmass = 3).
}\label{compar60}\end{figure}

We focussed our attention on the first diurnal sidelobe in the power spectrum.
The window function after 60 days of continuous observations is presented on  Fig.~\ref{compar60}. The energy around the day first alias at 11.57 $\mu$Hz is much reduced for Antarctica observation compared to network observation (Table~\ref{compalias}).

\begin{figure}
\plotone{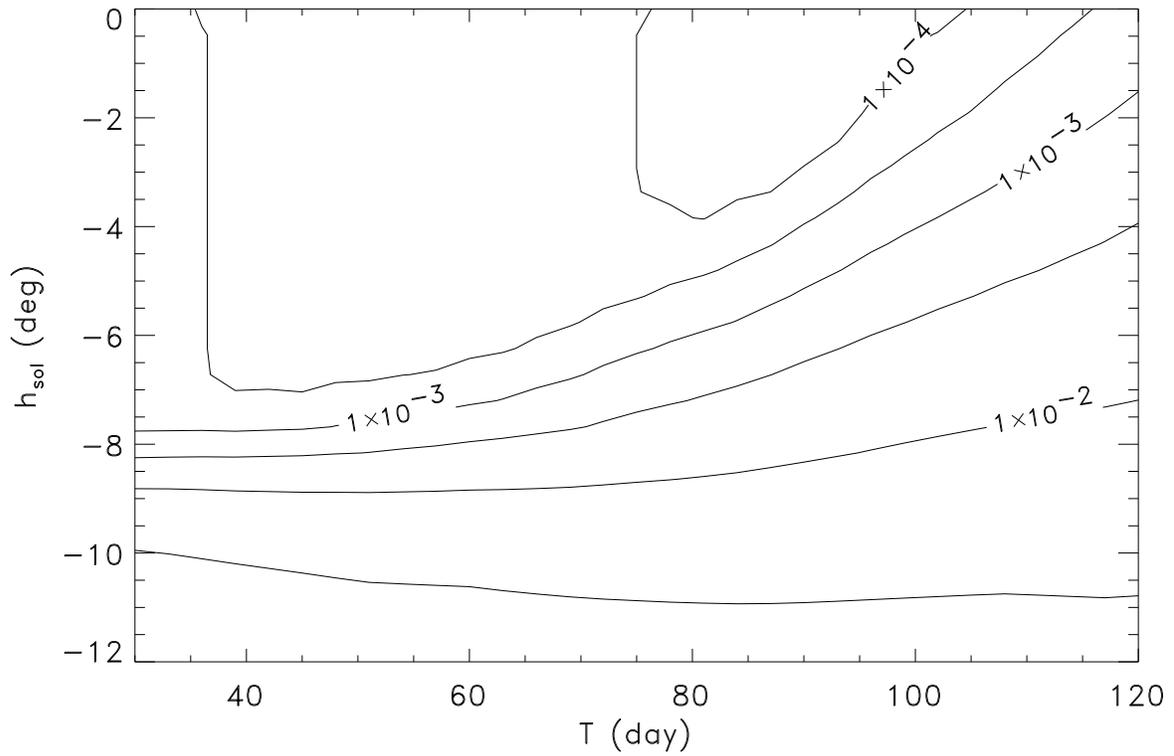}%\plotone{simuhsolduree.ps}
\caption{Contour plot of the mean energy around the first day
alias à 11.57 $\mu$Hz (Fig. \ref{compar60}), as a function of the
observation duration and maximum Sun altitude, derived from the
clear sky fraction observed in 2006. Performance is excellent for
bright targets observable as soon as the Sun is below  $-4^\circ$.
The energy in the daily alias increases with increasing
observation duration and with increasing restrictive condition on
the sky brightness.}\label{simuhsolduree}\end{figure}

The influence of the observation duration and Sun altitude below horizon can be discussed from Fig.~\ref{simuhsolduree}, that represents the relative mean energy around 11.57 $\mu$Hz, as a function of the observation duration and maximum Sun altitude. This contribution is about 5\% for faint targets requiring the Sun $-12^\circ$ below horizon. It lays in the range $10^{-3} - 10^{-4}$ for bright targets observable as soon as the Sun is below  $-4^\circ$. Observations in Antarctica with a first side lobe limited to 1\% of the signal may last as long as 4 months, providing therefore simultaneously excellent window and frequency resolution.

\subsection{Targets}

\begin{figure}
\plotone{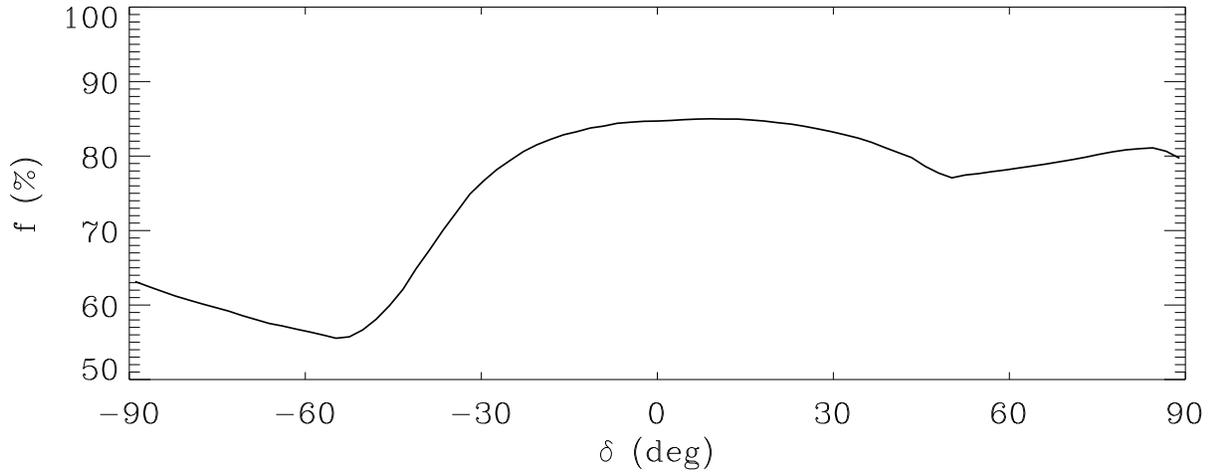}%\plotone{visudelta.ps}
\caption{Mean duty cycle for a 6-site network, as a
function of the declination $\delta$ of the target. The
asymmetrical shape of the curve relates the fact that the network
includes 4 sites in the Northern hemisphere, but only 2 in the
Southern hemisphere. Circumpolar stars benefit from their longer
visibility compared to lower declination targets, but are seen bad
by a too limited number of observatories of the network.
}\label{visudelta}\end{figure}

The observable sky fraction is limited to circumpolar targets at Dome C (Tab.~\ref{target}). Right ascensions opposite to the Sun are favored. The preparative work of the SIAMOIS project (Mosser et al. 2006) has identified the different targets and  proposed a scientific program for more than 6 winters. 
Conditions of network observations are quite different. The right ascension of the source plays no role, but the duty cycle varies as a function of declination in a complex way (Fig.~\ref{visudelta}). First,  for increasing absolute value of the declination, the duty cycle decreases due to asymmetry in the network and less efficiency of sites with latitude opposite to the stellar declination. Then, around respectively $+50^\circ$ and  $-50^\circ$, the decrease is hampered by the increase of visibility duration of high declinations sources, that compensates the low altitude above horizon. With 4 sites in the Northern hemisphere, but 2 in the Southern hemisphere, the duty cycle is lower for austral targets. Nevertheless, the band with a duty cycle above 80\% remains limited to declinations in the range $[-22^\circ, +35^\circ]$. This corresponds to 47\% of the sky, when only 29\% is observable at Dome C at airmass always less than 3.

\subsection{Number of sites}

As shown by the GONG site study, the number of sites is a crucial parameter for network observations. The performance of the network limited to 5 or 4 stations was therefore estimated. Simulations show that the star declination plays no important role, contrary to the identity of the absent observatories: the contribution of a Hawaiian site, without close neighbor, is of course highly valuable. The averaged duty cycle decrease is about 9\% with 5 sites, and 20\% with only 4 sites (Fig.~\ref{compar}). This decrease is slightly more pronounced with increasing observation duration. Note that a network with 3 sites in the same hemisphere may have performance comparable to a 4-site network, but with severe restrictions on the possible targets.

It is out of the scope of this paper to examine the question of maintenance, which is a crucial point for both network and Antarctica observations. At Dome C in Antarctica, the maintenance is limited to 1 site. Previous wintering (Agabi et al. 2006) has shown the potential of the site, with the logistic of the French and Italian polar institutes. Maintenance problems are also problematic for an efficient network, with an effort to be multiplied by the number of sites, in order to maintain all operational.
We did not take into account the Moon influence. At Dome C, the Moon will not interfere with the observations of circumpolar targets. For network observation, with targets at declination in the range $\pm 30^\circ$,  great care will have to be taken when selecting a target.

\section{Conclusion}

Observations conducted during the 2006 wintering at the Concordia station in Antarctica give an excellent clear sky fraction, that yields a meteorological  duty cycle as high as 92\%.  Consequently, asteroseismic observations conducted at Dome C during the 100-day long polar night provide performance better or similar than realized with a 6-site network. The only case where a 6-site network may provide better performance, but limited to a duty cycle about 77\%, is for a 5-month long observing run.
On the other side, asteroseismic observations at Dome C may provide a 100-day long time series with a total duty cycle about 87\%, what network observation cannot.
With such a duty cycle, the relative residual energy in the first diurnal sidelobes will be limited to less than $10^{-3}$ for bright targets ($mV\le 5$).
Performance on faint stars is about $10^{-2}$, limited by the sky brightness at noon local time.

\acknowledgments
Winterover of EA was granted by the French (Institut Paul Emile Victor) under the Concordiastro programme (PI E. Fossat, LUAN).
We thank the anonymous referee for his/her comments which helped clarifying the paper.

\clearpage

\begin{table}
\begin{tabular}{rrrrrrrrrrc} \hline
Sun limit $\hsun$&$0^\circ$&$-2^\circ$&$-4^\circ$&$-6^\circ$&$-8^\circ$&$-10^\circ$&$-12^\circ$\\ Duty cycle (\%)     &92.0     & 91.0     & 88.5     &  84.6    & 78.3     & 69.8      &    64.2 \\
\hline
\end{tabular}
\caption{Mean duty cycle at Dome C, over the 100-day long night,
as a function of the Sun altitude limit. Bright stars, that can be
observed without restrictive conditions on the Sun altitude, are
clearly favored.} \label{simuhsol}\end{table}

\begin{table}
\begin{tabular}{rrrrrrrr} \hline
target & type&$\delta$&  $V$ &$v\sin i$& A             \\
   &     &        &      & (km/s)  & (cm/s)\\
\hline
$\beta$Hyi  & G2IV&$-$77.25&  2.79&  5.0 &  59 \\
$\chi$Eri   & G5IV&$-$51.61&  3.71&  1.1 & 165 \\
  p Vel     & F4IV&$-$48.23&  3.84& 16.0 & 165 \\
HD114613    & G3V &$-$37.80&  4.85&   8.0&  73 \\
$\alpha$CenA & G2V&$-$60.83&$-$0.01& 2.7 &  34 \\
$\alpha$CenB & K1V&$-$60.84&  1.33&  1.1 &  18 \\
$\delta$Pav &G7IV &$-$66.18&  3.56&  3.2 & 111 \\
\hline
\end{tabular}
\caption{Possible targets list at Dome C, restricted to dwarfs and sub-dwarfs showing solar-like oscillations, with V magnitude brighter than 5, excluding giant or pulsators as $\delta$-Scuti. The oscillation maximum amplitude $A$ is estimated according to Samadi et al. (2005). All these targets can be observed with a 40-cm telescope.}
\label{target}\end{table}

\begin{table}
\begin{tabular}{lcccccccc} 
\hline
Duration& \multicolumn{4}{c}{Duty cycle}& \multicolumn{4}{c}{Energy at 11.6 $\mu$Hz}\\
(day)   & \multicolumn{4}{c}{(\%)}& \multicolumn{4}{c}{ (1 @ Dome C)}\\
         &Dome C & 6 sites & 5 sites & 4 sites &Dome C & 6 sites & 5 sites & 4 sites\\
\hline
40 & \sl94 & 84 & 76 & 68 & \sl1 & 5  & 15  & 40 \\
60 & \sl92 & 84 & 76 & 67 & \sl1 & 5  & 15  & 40 \\
80 & \sl90 & 83 & 75 & 66 & \sl1 & 1.3  & 5 & 20 \\
90 & \sl89 & 83 & 75 & 66 &    1 & \sl0.8  & 2 & 9\\
100& \sl87 & 82 & 74 & 65 &    1 & \sl0.5  & 1 & 4\\
\hline
\end{tabular}
\caption{Best duty cycle and energy of the window function in the frequency range around the day alias at 11.6 $\mu$Hz,  for different network configurations (4 $\to$ 6 sites) and different observation durations, normalized to the same energy obtained with the window function at Dome C. Best performance are highlighted in slanted text.  
With the energy criterion at 11.6 $\mu$Hz, the performance at Dome C is better only for observation durations less than 90 days. However, the duty cycle at Dome C remains greater than the one with a 6-site network for observation duration up to 120 days, so that the total energy distributed in the spectrum of the window function at non-zero frequency remains lower.}
\label{compalias}\end{table}

\end{document}